\let\latexdocument\document
\let\latexarabic\arabic

\documentclass[article,lineno]{biometrika}

\let\document\latexdocument
\let\arabic\latexarabic

\usepackage{amsmath}

\usepackage{times}
\usepackage{amssymb}
\usepackage{mathrsfs}  
\usepackage{bm}
\usepackage{natbib}
\usepackage{xcolor}
\usepackage{booktabs}
\usepackage{lipsum}

\newcommand\blfootnote[1]{%
  \begingroup
  \renewcommand\thefootnote{}\footnote{#1}%
  \addtocounter{footnote}{-1}%
  \endgroup
}

\newcommand{\bzero}{\boldsymbol{0}}
\newcommand{\btheta}{\boldsymbol{\theta}}

\newcommand{\bY}{\boldsymbol{Y}}
\newcommand{\bbQ}{\mathbf{Q}}
\newcommand{\bbSig}{\mathbf{\Sigma}}
\newcommand{\bbI}{\mathbf{I}}

\graphicspath{{./art/}}

\usepackage[plain,noend]{algorithm2e}

\makeatletter
\renewcommand{\algocf@captiontext}[2]{#1\algocf@typo. \AlCapFnt{}#2} 
\def\@algocf@capt@plain{top}
\renewcommand{\algocf@makecaption}[2]{%
  \addtolength{\hsize}{\algomargin}%
  \sbox\@tempboxa{\algocf@captiontext{#1}{#2}}%
  \ifdim\wd\@tempboxa >\hsize
    \hskip .5\algomargin%
    \parbox[t]{\hsize}{\algocf@captiontext{#1}{#2}}
  \else%
    \global\@minipagefalse%
    \hbox to\hsize{\box\@tempboxa}
  \fi%
  \addtolength{\hsize}{-\algomargin}%
}
\makeatother

\begin{document}



\markboth{Y. Zhang et~al.}{}

\title{Identification and Estimation of Heterogeneous Interference Effects under Unknown Network}

\author{Yuhua Zhang}
\affil{Department of Biostatistics, Harvard T.H. Chan School of Public Health
\email{yuhuazhang@hsph.harvard.edu}}

\author{Jukka-Pekka Onnela}
\affil{Department of Biostatistics, Harvard T.H. Chan School of Public Health
\email{onnela@hsph.harvard.edu}
}

\author{Shuo Sun$^*$}
\affil{Department of Population Health Sciences, Weill Cornell Medicine
\email{mis4060@med.cornell.edu}}

\author{\and Ruoyu Wang$^*$}
\affil{Department of Biostatistics, Harvard T.H. Chan School of Public Health
\email{ruoyuwang@hsph.harvard.edu}}

\maketitle

\begin{abstract}
Interference--in which a unit's outcome is affected by the treatment of other units--poses significant challenges for the identification and estimation of causal effects. Most existing methods for estimating interference effects assume that the interference networks are known. In many practical settings, this assumption is unrealistic as such networks are typically latent. To address this challenge, we propose a novel framework for identifying and estimating heterogeneous group-level interference effects without requiring a known interference network. Specifically, we assume a shared latent community structure between the observed network and the unknown interference network. We demonstrate that interference effects are identifiable if and only if group-level interference effects are heterogeneous, and we establish the consistency and asymptotic normality of the maximum likelihood estimator (MLE). To handle the intractable likelihood function and facilitate the computation, we propose a Bayesian implementation and show that the posterior concentrates around the MLE. A series of simulation studies demonstrate the effectiveness of the proposed method and its superior performance compared with competitors. We apply our proposed framework to the encounter data of stroke patients from the California Department of Healthcare Access and Information (HCAI) and evaluate the causal interference effects of certain intervention in one hospital on the outcomes of other hospitals.


\end{abstract}

\begin{keywords}
Causal Inference; Community Detection; Heterogeneous Treatment Effects; Interference Effects; Latent Interference Network.
\end{keywords}

\blfootnote{* Co-corresponding authors.}

\section{Introduction}

A key question in classical causal inference is the estimation of the average treatment effect (ATE) within the potential outcomes framework. For each unit $i$, $Y_i(1)$ and $Y_i(0)$ denote the outcomes that would be observed under treatment or control, respectively. Under random treatment assignment and standard assumptions, the difference in sample means is an unbiased estimator of the ATE. This framework relies on the Stable Unit Treatment Value Assumption (SUTVA), which includes (i) consistency and (ii) no interference—that is, each unit’s outcome depends only on its own treatment and not on the other units’ treatment assignments. However, in many real-world settings involving social, geographic, or organizational networks, interference is plausible: a unit’s outcome may depend on the treatments assigned to others. If interference is present but ignored, estimates targeting the ATE under the no interference assumption can be biased and make the estimate difficult to interpret. To address these challenges, alternative causal estimands and methods have been introduced that explicitly allow for interference, enabling identification and inference when treatments can affect multiple units
\citep{rosenbaum2007interference, hudgens2008toward, tchetgen2012causal, vanderweele2015interference}.


A key difficulty in the presence of interference is the explosion of the potential outcome space: each unit has a potential outcome for every combination of the other units’ treatment assignments. With binary treatment and $N$ total units, unit $i$ has $2^{N}$ potential outcomes. A common simplifying assumption is local (neighborhood) interference: a unit’s outcome depends on its own treatment and on the treatments of its network neighbors only. Under this assumption, many existing works study the estimation of population-level treatment effects \citep{manski2013identification, sussman2017elements, athey2018exact, hu2022average, ogburn2024causal}. In parallel, \cite{zhang2023individualized} develop a method to estimate heterogeneous group-level treatment effects. These approaches typically assume a known exposure mapping--a function that summarizes others’ treatments (e.g., the number or proportion of treated neighbors). A limitation is that identification can fail and estimators can be biased if the chosen mapping is misspecified, or if the observed network does not reflect the true interference network.


Recent work extends the study of network interference by allowing exposure mappings to be defined relative to a misspecified or noisy network. By \textit{misspecification}, we mean that some network edges are misclassified or, more generally, measured with error. Several papers study how sensitive causal conclusions are to such errors and propose estimators of causal effects that remain valid under specified forms of network misspecification \citep{leung2022causal, savje2024causal}. Other work treats the observed network as a noisy proxy for the true interference network and develops estimators that adjust for edge mismeasurement \citep{li2021causal}. However, these approaches typically require strong assumptions about how close the observed network is to the interference network (e.g., bounded or nondifferential misclassification) \citep{lewbel2024ignoring} or rely on simplified exposure mappings  \citep{li2021causal}, which can limit robustness and interpretability.


Another approach is to model the unknown interference network as random. \citet{li2022random} study settings in which the interference network is generated by the graphon model and define population-level indirect treatment effects, averaged over the network distribution. However, \citet{li2022random} assume the interference network is homogeneous in the sense that every pair of units has an equal probability of interference. This assumption may be unreasonable in many network settings. 


In this work, we establish identification conditions and estimation strategies for heterogeneous group-level indirect treatment effects when the interference network is random and unobserved. Assuming a latent group (community) structure in the network, we propose a Community Interference Network model (CINet) to estimate direct effect and group-level indirect (interference) effects. We define group-level causal estimands that characterize both within- and between-group interference effects and we allow these effects to vary across groups as a function of latent structural features of the network. Our approach proceeds in two steps. First, we identify the community structure in the observed network, assuming that the unobserved interference network shares the same latent communities but may differ in the probabilities of connection within and between communities. 
Second, we estimate the heterogeneous group-level effects and the within- and between-group interference probabilities conditional on the community labels. 
To address the intractability of the likelihood and to facilitate computation, we develop a Bayesian approach to estimate the causal effects and show that the posterior distribution concentrates around the MLE. 

Our contributions include: (i) developing a new causal inference framework with an unobserved heterogeneous interference network; (ii) establishing identification conditions for group-level interference effects, thereby establishing a new connection between the interference effect heterogeneity and identifiability; (iii) proposing an easy-to-implement estimator and establishing its asymptotic properties which enable valid inference; and (iv) showcasing the effectiveness of the proposed method through simulations and real data analysis.


The rest of the paper is structured as follows: Section 2 introduces notation and defines our estimands: the average direct causal effect and group-level indirect effects. Section 3 presents the model, establishes identification conditions, and derives the asymptotic properties of the proposed estimator. We demonstrate our method through simulations in Section 4 and apply the method to the stroke patient encounter data from California Department of Health Care Access and Information (HCAI) in Section 5. Section 6 concludes the paper with a discussion.





\section{Preliminaries}

\subsection{Motivating Example}
\label{subsec:mot_exp}

Our motivating example is acute ischemic stroke, a leading cause of long-term disability in older population. Timely reperfusion therapies can substantially improve outcomes, but their effectiveness depends on access within a narrow treatment window. Endovascular Thrombectomy (EVT)—a minimally invasive procedure that removes blood clots from cerebral arteries—has been shown to significantly improve recovery in eligible patients. In practice, individuals typically present first to the nearest hospital, many of which lack EVT capability. Access is therefore often achieved through emergency transfer to specialized stroke centers. This structure implies that EVT availability at one hospital can benefit not only its own patients but also those initially presenting to surrounding hospitals, highlighting the potential for interference effects across the healthcare network.

Modeling these interference effects within a causal framework introduces key challenges. The observed transfer network may not represent the true exposure network, since not all transfers are related to EVT. Moreover, hospitals often form clusters, with transfers occurring more frequently within clusters than between them, suggesting that hospitals in the same cluster have stronger influence on one another. These features motivate our central question: how can we identify and estimate causal effects when the interference network is unobserved but exhibits latent community structure?

\subsection{Setup}
\label{subsec:prelim}
Suppose one has observed data on a population of $N$ interconnected units. For each $i\in \{1,...,N\}$, we observe the treatment (or intervention) $Z_i \in \{0,1\}$ and the corresponding outcome $Y_i$. Let $\boldsymbol{Z}=(Z_1,...,Z_N)\in \{0,1\}^{N}$ denote the treatment assignment vector for the entire population. Let the potential outcome for the entire population be $\boldsymbol Y(\boldsymbol z)=\{Y_1(\boldsymbol z),...,Y_N(\boldsymbol z)\}$, where $Y_i(\boldsymbol z)$ denote the  potential outcome indexed by the treatment allocation vector $\boldsymbol{z} \in \mathcal{Z}(N)$, where $\mathcal{Z}(N)$ is the set of all possible treatment allocations of length $N$. Under this binary treatment setting, the cardinality of the set $\mathcal{Z}(N)$ is $2^N$. We consider a randomized controlled trial (RCT) and impose the following standard causal assumptions: (i) consistency: $\bY(\boldsymbol{Z})=\bY$ if the realized treatment assignment is $\boldsymbol{Z} = \boldsymbol{z}$; (ii) positivity: every treatment allocation of interest has a positive probability, $P(\boldsymbol{Z} = \boldsymbol{z}) > 0$ for all $\boldsymbol{z} \in \mathcal{Z}(N)$. Randomization ensures exchangeability by design.

We use a structural equation model to unify the causal effects of interest with the network structure. Specifically, we assume the data are generated by the following model:
\begin{align}\label{eq:SEM}
    Y_i=f(\mathbf{G},\boldsymbol{Z})+\varepsilon_i, \quad i=1,...,N,
\end{align}
where $f$ is an unspecified function, $\mathbf{G}$ is the latent interference network, and $\varepsilon_i$ is an unobserved error for unit $i$. This specification indicates that the outcome $Y_i$ depends on both the treatment assignments and the exposure mapping for each unit $i$. We assume $\varepsilon_1,...\varepsilon_N$ are independent and identically distributed (iid).

Next, we specialize the setup to the network setting.  Let $\mathbf{A} = [A_{ij}]$ denote the adjacency matrix of a network $\mathcal{N} = (V, E)$, where $V$ is the set of nodes and $E$ the set of edges. For a population of $N$ interconnected units, each unit corresponds to a node, so that $|V| = N$. The network is represented by an $N \times N$ adjacency matrix $\mathbf{A}$, with entries defined as $A_{ij} = 1$ if there is an edge from node $i$ to node $j$, and $A_{ij} = 0$ otherwise. We distinguish between the observed network $\mathcal{N} = (V, E)$ with adjacency matrix $\mathbf{A}$, and the latent interference network $\mathcal{N}'=(V',E')$ with adjacency matrix $\mathbf{G}=[G_{ij}]$, where $G_{ij}=1$ indicates the existence of interference effect between unit $i$ and $j$. We assume the network is directed and exclude the self-loops.

In general, $\mathbf{A}\neq \mathbf{G}$, and the interference network $\mathbf{G}$ is unknown. Most existing methods assume $\mathbf{A} = \mathbf{G}$; we relax this assumption, allowing for discrepancies between the two and explicitly accounting for uncertainty in the inteference network. Details of this formulation are provided in the Methods section. Throughout the paper, for any collection $\{X_i\}_{i=1}^N$, we use $\mathbf{X}=(X_1,...,X_N)$ to denote the complete vector and $\mathbf{X}_{-i}=(X_1,...,X_N)\backslash X_i$ to denote the vector excluding the $i$th element.


\subsection{Causal Estimand}

The parameters of interest are the average direct causal effect (DE) and the group-level average indirect causal effects (IDE). The sample DE is given by
$$\tau_{\text{DE}}(\boldsymbol{z})=\frac{1}{N}\sum_{i=1}^{N}E[Y_i(z_i=1, \boldsymbol{z}_{-i})-Y_i(z_i=0, \boldsymbol{z}_{-i})].$$
This estimand represents the average causal effect of a unit's own treatment on its own outcome, holding the treatment assignments of all other units fixed at $\boldsymbol{z}_{-i}$. It represents the direct contribution of a unit's treatment apart from any influence of other units' treatment assignments.

We assume the interference is directed; that is, the interference is initiated by the sender and affects the receiver. Denote $V'_{l}\subseteq V',\ l\in\mathbb{N}$ as a subset of all units 
and $N_l=|V'_l|$ as the number of units in $V'_l$. Denote $V'_s\subseteq V'$ as the set of senders and $V'_r\subseteq V'$ as the set of receivers. We define the group-level IDE as 
$$\tau_{\text{IDE}}(V'_s, V'_r,\boldsymbol{z})=\frac{1}{N_{r}}\sum_{i\in V'_r}\sum_{j\in V'_s\backslash \{i\}}E[Y_i(z_j=1, \boldsymbol{z}_{-j})-Y_i(z_j=0,\boldsymbol{z}_{-j})].$$ 
The $\tau_{\text{IDE}}(V'_s, V'_r,\boldsymbol{z})$ represents the average change in outcomes for units in the receiver group $V'_r$ when the treatment status of units in the sender group $V'_s$ switches from $0$ to $1$, holding all other treatments fixed at $\boldsymbol{z}_{-j}$. By averaging across all receivers in $V'_r$ and all senders in $V'_s$, it captures the interference effect transmitted from one group to another.

Two special cases illustrate the definition. First, when $|V'_s| = |V'_r| = 1$, the estimand reduces to the pairwise indirect effect, measuring the interference of a sender on a receiver. Second, when $V'_s = V'_r = V'$, the estimand becomes $(1/N)\sum_{i=1}^{N}\sum_{j\neq i}E[Y_i(z_j=1, \boldsymbol{z}_{-j})]-E[Y_i(z_j=0,\boldsymbol{z}_{-j})]$, which is the average indirect effect defined in ~\citet{hu2022average}. Our formulation thus provides a general framework that enables estimating IDEs between any arbitrary groups.





\section{Method}

\subsection{Model Specification}

In this section, we specify a linear version of the structural equation model introduced in Eq.~\eqref{eq:SEM}. Specifically, we consider the model
\begin{align}
\label{eq:model1}
Y_i=\beta_0+\sum_{j=1,j\neq i}^N\beta_{ji}G_{ji}Z_j+\gamma Z_i+\varepsilon_i,
\end{align}
where $\gamma $ denotes the DE of unit $i$'s own treatment on its outcome and $\beta_{ji}$ captures the size of the interference effect size of unit $j$'s treatment on unit $i$'s outcome, transmitted through the interference network edge $G_{ji}$.

The true interference network $\mathbf{G}$ is not observed, and the observed network $\mathbf{A}$ provides only partial information about it. To proceed, we assume
the node sets coincide (i.e., $V=V'$) and posit a relationship between $\mathbf{G}$ and $\mathbf{A}$ by assuming a shared latent cluster structure in both networks. This assumption is less restrictive than requiring $\mathbf{G}=\mathbf{A}$, since it allows for discrepancies at the edge level, and it is more flexible than models that rely on a prespecified simplified exposure mapping, which may fail under misspecification. In the stroke transfer setting, hospitals naturally cluster by geography, affliations, and service capacity; even if observed patient transfers in $\mathbf{A}$ do not indicate existence of interference, hospitals within the same cluster tend to exhibit similar interaction patterns and have comparable influence on one another, making a shared community structure a plausible bridge from $\mathbf{A}$ to the latent $\mathbf{G}$. 

We perform community detection on the observed network $\mathbf{A}$ to identify the cluster labels, which can be formulated as finding a disjoint partition $V=V_1\cup...\cup V_K$ of nodes, and each node set is called a community. Denote $\boldsymbol{C}=\{C_1,...,C_N\}$ as the node labels, where $C_i$ is the label of node $i$ and takes values in $\{1,...,K\}$. The community detection problem is equivalent to identifying the latent labels $\boldsymbol{C}$. A widely accepted model in community detection is the stochastic block model (SBM), which assumes that, given the node labels $\boldsymbol{C}$, the entries of the adjacency matrix $\mathbf{A}$ are independent Bernoulli random variables with $E(A_{ij}\mid C_{i}, C_{j})=P_{C_iC_j}$. The collection $\mathbf{P}=[P_{ab}]_{a,b=1}^K$ is a $K\times K$ matrix of within- and between-community connection probabilities. 
Associated with SBM is an objective function, enables the latent community labels to be inferred. Common choices include the Erdős–Rényi modularity and the likelihood function of the SBM. 
The strong consistency of community label estimation, that is, $P(\hat{\boldsymbol{C}}=\boldsymbol{C})\rightarrow 1,\ \text{as\ }N\rightarrow \infty$, has been established for the SBM under certain objective functions by~\citet{zhao2012consistency}:
\begin{proposition}\label{prop}
In the stochastic block model, the model criterion (e.g., likelihood and Erdős–Rényi modularity) is consistent when $\lambda_N/ \log N\rightarrow \infty$ as $N\rightarrow\infty$, where $\lambda_N$ is the expected degree of the graph.
\end{proposition}

Proposition~\ref{prop} was establish for the observed network, here denoted by $\mathbf{A}$. We specify the connection between the community structure in $\mathbf{A}$ and that in $\mathbf{G}$ formally.
Denote $\boldsymbol{C}_{\rm A}$ and  $\boldsymbol{C}_{\rm G}$ as the two sets of community labels in $\mathbf{A}$ and $\mathbf{G}$ respectively, each taking values in $[K]^{N}$. We assume that $\mathbf{A}$ and $\mathbf{G}$ share the same set of community labels.

\begin{assumption}\label{assum:rho}
(Shared community labels).
 There exists a bijection $\rho$ such that $\boldsymbol{C}_{\rm A}=\rho\boldsymbol{C}_{\rm G}$. That is, the difference between $\boldsymbol{C}_{\rm A}$ and $\boldsymbol{C}_{\rm G}$ is up to a permutation $\rho$. 
\end{assumption}

Assumption~\ref{assum:rho} indicates that Proporsition~\ref{prop} can be extended to $\mathbf{G}$. In other words, if the community labels $\mathbf{C}_A$ are consistent, then the community labels $\mathbf{C}_G$ are consistent. We say that nodes share the same labels in $G$ form \textit{communities} as well.
Similar to that in the SBM, we assume nodes in the same community have the same probability of connecting with other nodes in the interference network. 
Formally, 
\begin{assumption}\label{assum:pi}
(Community structure in the interference network).
 Let $\pi_{kk'}=P(G_{ji}=1|C_i=k,C_j=k')$, where $\pi_{kk'}$ denotes the probability of an edge between any node in community $k$ and any node in community $k'$.   
\end{assumption}
By analogy, we assume community-level interference effects, meaning that the interference effect size between two units is determined solely by their community labels:
\begin{assumption}\label{assum:group-level-beta}
(Community-level interference effects). 
The interference coefficient depends only on the community labels of the sender and receiver units. 
Formally, $\beta_{ji} = \beta_{j'i'}$ whenever $C_j = C_{j'}$ and $C_i = C_{i'}$.
\end{assumption}
We propose a two-step estimation procedure. In the first step, we infer the community labels of each unit by performing community detection on the observed network. In the second step, conditional on the estimated labels, we estimate the model parameters, including DE and group-level IDEs. Given that our main interest focuses on the second step, all subsequent discussion is conducted conditional on the community labels $\boldsymbol{C}$. Denote the group-level interference effect size of the $k{\text{th}}$ group on the $k'{\text{th}}$ group as $\beta_{k,k'}$. Given $C_i=k^{\prime}$ and under Assumption~\ref{assum:group-level-beta}, we can rewrite the linear model in Eq.~\eqref{eq:model1} as: 
\begin{align}\label{eq:mod2}
Y_i=\beta_0 + \sum_{k=1}^K \beta_{k,k^{\prime}} \sum_{j=1, C_j=k, i\neq j}^{N_k}G_{ji}Z_j + \gamma Z_i +\varepsilon_i, 
\end{align}
where $N_k$ is the number of nodes in the $k{\text{th}}$ community.
Compared to other methods, our proposed model requires neither prior knowledge of the unknown exposure mapping $\mathbf{G}$ nor strong assumptions on the similarity between $\mathbf{A}$ and $\mathbf{G}$ as discussed in~\citet{lewbel2024ignoring}. Moreover, our proposed model allows for group-wise heterogeneous interference effects, which avoids over-simplifying the complex nature of the interference network and maintains flexibility compared to other methods~\citep{li2021causal}. 

\subsection{Identifiability and Estimation}
Identifiability is fundamental in causal inference, as it determines whether the causal effects can be uniquely and reliably recovered from observed data distribution. In this section, we will show that the parameters in Eq.~\eqref{eq:mod2} can be uniquely identified once the latent community labels $\boldsymbol{C}$ are recovered with strong consistency and the interference effect sizes exhibit sufficient heterogeneity across communities. To make this construction explicit, let $C_i=k^{\prime}$; the linear specification in Eq.~\eqref{eq:mod2} can now be written as
\begin{align}\label{eq:outcome_mod_Q}
    Y_i=\beta_0 + \sum_{k=1}^K \beta_{k,k^{\prime}} Q_{k,i} + \gamma Z_i +\varepsilon_i,
\end{align}
where $Q_{k,i}=\sum_{j=1, C_j=k, i\neq j}^{N_k}G_{ji}Z_j$ is the aggregated treatment exposure that unit $i$ receives from units in community $k$ (excluding itself if $k'=k$), and iid $\varepsilon\sim N(0, \sigma_{\varepsilon}^2)$. We impose the following assumptions:
\begin{assumption}\label{assm:z_indept}
(Independence and exchangeability of edges).
For any distinct $\{j,i\}\neq \{l,m\}, j\neq i, l\neq m$, we have $G_{ji} \perp  G_{lm} | \boldsymbol{C}$. And if $C_{i}=C_{i'}$, $G_{ji}$ and $G_{ji^{\prime}}$ are identically distributed conditional on  $\boldsymbol{C}$. 
\end{assumption}
\begin{assumption}\label{assm:identification}
(Heterogeneous interference effect sizes).
For any $k_{1}\neq k_{2}$, $k_{1}, k_{2}, k\in [K]$, the interference effect sizes differ across sender communities: $\beta_{k_{1},k}\neq \beta_{k_{2},k}$.
\end{assumption}
\begin{assumption}\label{assm:nk}
(Proportional community size ).
For all $k\in [K]$, $N_{k}/N\rightarrow\rho_{k}$ for some constant $\rho_{k} > 0$ almost surely as $N\rightarrow\infty$. 
\end{assumption}
\begin{assumption}\label{assm:n_pi}
(Finite expected number of senders).
For all $k,k'\in[K]$, assume that $\pi_{k,k'}\in(0,\delta]$ for some constant $\delta<1$ and $n_{k}\pi_{k, k'}\rightarrow \lambda_{k,k'}<\infty$ almost surely as $N\rightarrow\infty$, where $n_{k}= \sum_{j=1, C_j=k}^{n} Z_j$ is the number of treated units in community $k$ excluding unit $i$.
\end{assumption}

Assumption~\ref{assm:z_indept} specifies that conditional on the latent community labels, edges are independent across node pairs, and nodes within the same community are exchangeable in terms of how they connect to others. Under this assumption, $Q_{k,i} \mid C_i,\{Z_j\}_{j\neq i} \sim \mathrm{Binom}(n_{k},\ \pi_{k,C_i})$. 
Assumption~\ref{assm:identification} requires heterogeneous group-level interference effect sizes across communities, which is essential to establish the identifiability (Theorem~\ref{thm:indentification}) below. Intuitively, if two communities send identical interference effects on a given receiver community, their contributions cannot be distinguished from observed outcomes. Assumption~\ref{assm:nk} is a regularity condition that prevents degenerate cases where some communities become negligible asymptotically. It ensures that each community occupies a non-negligible fraction of the population, which is necessary for consistent estimation of community-level parameters as the network grows. 

Assumption~\ref{assm:n_pi} imposes finite number of expected senders on the interference network. The bound $\pi_{kk'} \in (0,\delta]$ prevents communities from forming fully connected graphs, while the scaling condition $n_k \pi_{kk'} \to \lambda_{kk'} < \infty$ ensures that the expected number of within- and between-community connections remains finite as the network grows. Since the size $n_k $ grows with $N$, this requires the individual-level interference probability $\pi_{kk'}$ to diminish toward zero.

Denote the parameter vector $\btheta_N=\{\{\beta_{k,k'}: k,k'\in[K]\},\{\lambda_{k,k'}^{(N)}: k,k'\in[K]\},\sigma_{\varepsilon}, \beta_0, \gamma \}$, where $\lambda_{k,k'}^{(N)}=n_k\pi_{k,k'}$. To study the identification in the limit where $N \to \infty$, define 
\begin{align}\label{eq:f_Y_pois}
    \widetilde{Y}_i=\beta_0 + \sum_{k=1}^K \beta_{k,k'} \widetilde{Q}_{k,i} + \beta_{\rm{dir}}Z_i +\varepsilon_i,
\end{align}
where the Binomial term $Q_{k,i}$ in Eq.~\eqref{eq:outcome_mod_Q} is replaced by a Poisson variable $\widetilde{Q}_{k,i} \sim \mathrm{Pois}(\lambda_{k,k'})$. This corresponds to the limiting distribution of the Binomial variable $Q_{k,i}$ given $C_i, \{Z_j\}_{j\neq i}$ as $N\rightarrow\infty$ under Assumption~\ref{assm:n_pi}, which we will use to establish the identification theorem. 
Notably, for any $i, j$ such that $C_i = C_j$ and $Z_i = Z_j$, the likelihood function for $\widetilde{Y}_i$ and $\widetilde{Y}_j$ are identical. Consequently, the observed data identify $2 K$ different conditional distributions of $\widetilde{Y}$, i.e., the distribution of $\widetilde{Y}$ conditional on $C = c, Z = z$ for $c\in [K]$ and $z\in \{0, 1\}$. Denote the corresponding conditional density as $p(\tilde{y}\mid c, z; \btheta_{\infty})$, where $\btheta_{\infty} = \lim_{N\to \infty} \btheta_N = \{\{\beta_{k,k'}: k,k'\in[K]\},\{\lambda_{k,k'}: k,k'\in[K]\},\sigma_{\varepsilon}, \beta_0, \gamma \}$. To formalize identifiability in this setting, we define it as follows. 
\begin{definition}\label{def:def}
    The limiting parameter $\btheta_{\infty}$ is said to be identifiable if: for any $\btheta^{\prime}$ in the parameter space, $p(\tilde{y}\mid  c, z; \btheta_{\infty}) = p(\tilde{y}\mid  c, z; \btheta^{\prime})$ for $c\in [K]$ and $z\in \{0, 1\}$ only if $\btheta^{\prime} = \btheta_{\infty}$ up to some permutation of the community labels.
\end{definition}


Since the distributions of $Z_i$ and $C_i$ are irrelevant to $\btheta_{\infty}$, in the limit, the functions $p(\tilde{y}\mid c, z; \btheta_{\infty})$ contain all the information in the observed data about the parameter $\btheta_{\infty}$. Building on Definition~\ref{def:def},  we now characterize the precise condition under which it holds in our model.
\begin{theorem}\label{thm:indentification}
(Identifiability):  The limiting parameter $\btheta_{\infty}$ is identifiable if and only if Assumption \ref{assm:identification} holds.    
\end{theorem}
Theorem~\ref{thm:indentification} shows that heterogeneity in group-level interference effects across communities is both necessary and sufficient for identifiability, and thus for the consistency of the MLE (Section~\ref{subsec:mle}). If two sender communities exert identical interference effects on the same receiver community, their contributions are indistinguishable in the likelihood, preventing unique recovery of the true parameter. From an applied standpoint, it means that causal effects of interventions under interference can be validly estimated and interpreted even when the interference network is unobserved, provided that community-level heterogeneity exists.



\subsection{Asymptotic Properties of MLE}\label{subsec:mle}
Under Assumption \ref{assm:z_indept}, and conditional on community labels $\boldsymbol{C}$ and treatments $\boldsymbol{Z}$, the likelihood of the observed outcomes $\bY = (Y_1, \dots, Y_N)^{\top}$ is
\begin{align}
\label{eq:llk1}
\begin{split}
\small
p(\bY\mid \boldsymbol{C}, \boldsymbol{Z}; \btheta_N)&=
\prod_{i=1}^{N}\sum_{k = 1}^{K}\sum_{q_{k, i} = 1}^{n_{k} - \Delta_{i, k}}\frac{1}{\sqrt{2\pi\sigma_{\varepsilon}^2}}\exp\left\{-\frac{(Y_i-\beta_0-\sum_{k=1}^{K}\beta_{k,C_i}q_{k,i}-\gamma Z_i)^2}{2\sigma_{\varepsilon}^2}\right\}\times\\
&\quad\prod_{k=1}^{K}\left(^{n_k - \Delta_{i, k}}_{\quad q_{k,i}}\right)\pi_{k,C_i}^{q_{k,i}}(1-\pi_{k,C_i})^{n_k - \Delta_{i, k} - q_{k,i}},
\end{split}
\end{align}
where $\Delta_{i, k} = 1\{C_i = k\}Z_i$. Let $l_N(\btheta_N) = N^{-1}\log p(\bY\mid \boldsymbol{C}, \boldsymbol{Z}; \btheta_N)$ be the normalized log conditional likelihood, and $\hat{\btheta}_N$ the MLE that maximizes $l_N(\btheta_N)$.
In this section, we discuss the asymptotic properties of the MLE. 
The following theorem shows that the MLE $\hat{\btheta}_N$ converges to $\btheta_N$.

\begin{theorem}\label{thm:MLE}
(Consistency of MLE): 
Under Assumptions 1-4. The MLE estimator $\hat{\btheta}_N$ is consistent in the sense that 
for every $\epsilon>0$,
$$P(\|\hat{\btheta}_N-\btheta_N\|\ge\epsilon)\rightarrow 0.$$
\end{theorem}

Note that the likelihood given by Eq.~\ref{eq:llk1} changes with the sample size $N$, which is one challenge in proving Theorem~\ref{thm:MLE}. By connecting the Binomial distribution with Poisson, we are able to prove the consistency of the MLE. One interpretation is that the MLE of parameters are consistent if the combined interference effects from each group converge to a constant as $N\rightarrow\infty$.

\begin{assumption}
    \label{assmp:eigen}
    Let $\lambda_{{\rm min}, N}$ be the minimal eigenvalue of the population information matrix at $\btheta_N$.
    There exists a positive number $M_1>0$ such that
     $$\lambda_{{\rm min}, N}> M_1.$$
\end{assumption}

We show in the Appendix via simulation that the empirical minimal eigenvalues are bounded away from 0 as $N\rightarrow \infty$. Assumption~\ref{assmp:eigen} is essential to prove Theorem~\ref{thm:asynorm}, which shows the limiting Gaussian distribution of the MLE.

\begin{theorem}\label{thm:asynorm}
(Asymptotic Normality): Under Assumptions 1-5. Let $\bbSig_{k}=-E[\nabla^2_{\btheta}l_N(Y_i;\btheta)\mid C_{i} = k]$ for $k = 1,\dots, K$ and $\bbSig=\frac{1}{N}\sum_{k=1}^{K}N_k\bbSig_{k}$, where $\nabla_{\btheta}$ denotes the partial derivative.  
Then, we have
$$\sqrt{N}\bbSig^{-1/2}(\hat{\btheta}_N-\btheta_N)\rightarrow N(\bzero,\bbI)\text{ in distribution}.$$
\end{theorem} 

Although the form of the MLE is straightforward, in practice, the numerical calculation can be computationally intractable. Here, the Bayesian framework provides a solution. The Bayesian posterior also provides additional distributional information compared with the MLE.

\subsection{Bayesian Estimator}

We implement a Gibbs sampling algorithm to infer the posterior of model parameters $\btheta_N=\{\{\beta_{k,k'}: k,k'\in[K]\},\{\lambda_{k,k'}^{(N)}: k,k'\in[K]\},\sigma_{\varepsilon},\gamma \}$. We then show the asymptotic properties of the posterior by relating it with the MLE. The complete likelihood of the model is given in Eq.~\ref{eq:llk1}.
Let $\bbQ = (Q_{k, i})_{k\in [K], i \in [N]}$.
We treat the variables $\bbQ$ as latent variables, such that the joint likelihood of $\bY$ and $\bbQ$ is given by
$$
\begin{aligned}
    p(\bY,\bbQ\mid\btheta)=&
    \prod_{i=1}^{N}\frac{1}{\sqrt{2\pi\sigma_{\varepsilon}^2}}\exp\left\{-\frac{(Y_i-\beta_0-\sum_{k=1}^{K}\beta_{k,C_i}Q_{k,i}-\gamma Z_i)^2}{2\sigma_{\varepsilon}^2})\right\}\\
    &\prod_{k=1}^{K}\left(^{n_k - \Delta_{i, k}}_{\quad Q_{k,i}}\right)\pi_{k,C_i}^{Q_{k,i}}(1-\pi_{k,C_i})^{n_k - \Delta_{i, k} - Q_{k,i}}.
\end{aligned}
$$
The posterior of the model parameters is given by $p(\btheta|\bY,\bbQ)\propto p(\bY,\bbQ|\btheta)p(\btheta),$ where $\btheta$ is the model parameter and $p(\btheta)$ is the prior. We put the following assumption on the priors: 

\begin{assumption}
\label{assmp:prior}
    There exists a real number $\delta>0$ the same as in Assumption~\ref{assmp:eigen} and a real number $M_2>0$, such that
    $$\sup_{\btheta:||\btheta-\btheta_N||\le \delta}\left|\frac{1}{p(\btheta)}\right|\le M_2.$$
\end{assumption}

Assumption~\ref{assmp:prior} states that the prior puts a non-negligible weight in the $\delta$-neighbourhood of the MLE. We further establish the asymptotic properties of the posterior.

\begin{theorem}\label{thm:bvm}
(Bernstein von Mises Theorem):  Under Assumptions 1-6. Let the $TV$ denote the total variation distance. For the posterior estimator $\tilde{\btheta}_n$, we have the following conclusion:
$$TV(\mathcal{L}(\sqrt{N}(\tilde{\btheta}_N-\hat{\btheta}_N)),N(0,\hat{J}_N(\hat{\btheta}_N)^{-1}))\rightarrow 0.$$

\end{theorem}

Note that our proposed estimands of causal effects under the linear model specification are linear combinations of the parameters. Therefore, the consistency and asymptotic properties are generalizable to the estimands of causal effects.

\subsection{Justification on Covariates}

So far, we have assumed the treatment assignment has been randomized. In practice, we need to adjust for covariates to refine the analysis of the treatment effects by taking into account the fact that some baseline characteristics are related to the outcome and may be unbalanced between treatment groups. Failure to adjust for certain covariates will lead to inefficient or biased estimation.
We assume the covariates are independent of treatment assignments and the interference network: $\mathbf{X}\perp \mathbf{G}$, and $\mathbf{X}\perp \boldsymbol{Z}$.
Given $C_{i} = k^{\prime}$, the linear model adjusted for covariates can be written as
\begin{align*}Y_i=\beta_0 + \sum_{k=1}^K \beta_{k,k^{\prime}} \sum_{j=1, C_j=k, i\neq j}^{n_k}G_{ji}Z_j + \gamma Z_i +\beta_{X}X_i+\varepsilon_i,\end{align*}
where $\beta_X$ represent the coefficients of covariates $\mathbf{X}$.

\section{Simulation}

\begin{figure}
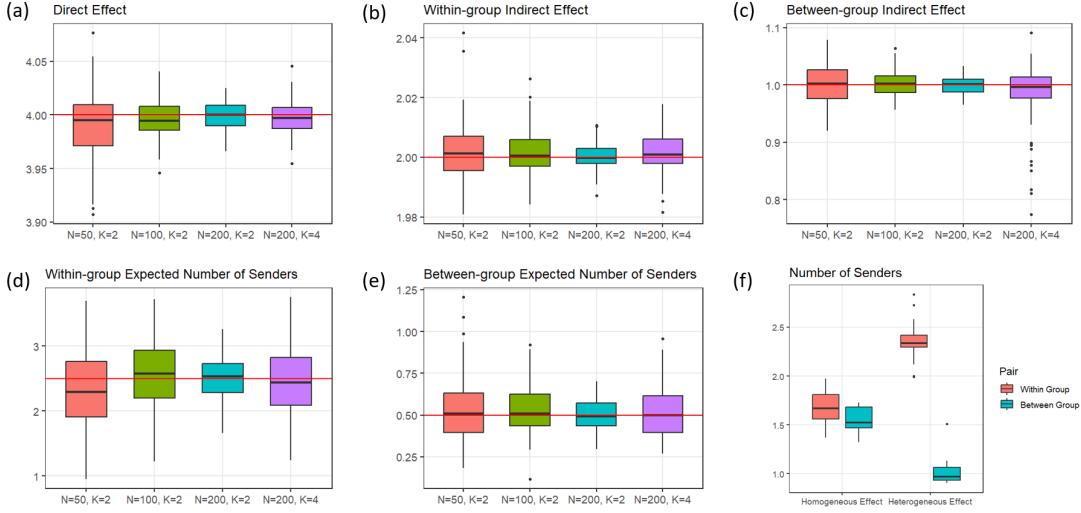

\figuresize{.3}
\figurebox{20pc}{25pc}{}[fig1]
\caption{Boxplots of (a) direct effect $\gamma $, (b) and (c) indirect effects $\beta_{k,k'}$, and (d) and (e) expected number of senders $\lambda_{k,k'}^{(n)}$. Redlines indicate the true values. (f) Boxplots of within- and between-group expected number of senders given homogeneous and heterogeneous indirect treatment effects.}
\label{fig1}
\end{figure}

To validate our findings from Theorem~\ref{thm:MLE},~\ref{thm:asynorm}, and~\ref{thm:bvm}, we consider a numerical example. We simulate different settings with the sample size set to $N=50,\ 100,\ 200$, and the number of communities set to $K=2, 4$ respectively. For a network with $N$ nodes, the adjacency matrix of $\mathbf{A}$ is generated by the stochastic block model given the number of community $K$, with the within-block connection probability equal to 0.5 and the between-block connection probability equal to 0.1. We generate the block assignment of each node i.i.d. from a categorical distribution with equal probability $1/K$ for any of the $K$ blocks. We then generate the interference network given the same community labels with the within block connection probability $\pi_1$ and the between block connection probability $\pi_2$. Given $N=100$ and $K=2$, we let $\pi_1=0.1$ and $\pi_2=0.02$. In other settings, $\pi_1$ and $\pi_2$ are chosen such that $\lambda_{k,k'}^{(N)}=n_{k}\pi_{k,k'}$ stays the same. We simulate the outcome data according to Eq.~\ref{eq:model1} with $\varepsilon_i\sim N(0,\sigma_{\varepsilon}^2)$. We generate the treatment assignments as $Z_i\sim \text{Bernoulli(0.5)}$ for each unit $i\in [N]$. The direct treatment effect is set to be $\gamma =4.0$, while the indirect treatment effects are set to be $\beta_{k,k'}$. When $K=2$, we let $\beta_{k,k'}=2.0$ if $k'=k$, and $\beta_{k,k'}=1.0$ if $k'\neq k$. When $K=4$, we let $\beta_{k,k'}=2.0$ if $k'=k$, and $\beta_{k,k'}=0.5, 1.0, 1.5$ if $k'\neq k$. We then apply the proposed two-step estimation procedure to the simulated data. In the first step, we infer the community labels given the adjacency matrix $\mathbf{A}$. In the second step, we estimate the direct causal effect and group-wise indirect causal effects given the inferred community labels. Figure~\ref{fig1} shows the posterior distributions of the direct causal effects $\gamma $, the indirect causal effects $\beta_{k,k'}$, and expected number of senders $\lambda_{k,k'}^{(N)}=n_k\pi_{k,k'}$ in different settings with 100 repeats. We see the posterior samples of the estimators center around the true values and closely match the limiting Gaussian distribution from Theorem~\ref{thm:asynorm}.  

We next compare our proposed method with other existing methods. Specifically, we compare our proposed method to the Horvitz–Thompson estimator of the direct causal effect~\citep{horvitz1952generalization}, the average direct and indirect causal effects under fixed interference network~\citep{hu2022average}, and the average direct and indirect causal effects under random network~\citep{li2022random}. Table~\ref{table1} shows the mean and standard deviation of the estimators across 100 repeated simulations. For a fair comparison with other methods, we report the population-average indirect treatment effect, defined (for a given receiver group $k'$) as $\sum_{k}\tilde{\beta}_{k,k'}n_k\tilde{\pi}_{k,k'}$, where $\tilde{\beta}_{k,k'}$ and $\tilde{\pi}_{k,k'}$ are posterior means.

\begin{table}
\caption{Mean and standard deviation of the direct and indirect causal effect estimators.}
\resizebox{\textwidth}{!}{\begin{tabular}{lcccccc}
\toprule
& \multicolumn{2}{c}{$N=100, K=2$}& \multicolumn{2}{c}{$N=200, K=2$} & \multicolumn{2}{c}{$N=200, K=4$} \\
& Direct & Indirect & Direct & Indirect & Direct & Indirect \\[5pt]
\midrule
IPW & 4.09 (0.67) & -- & 4.05 (0.44) & -- & 4.08 (0.34)& -- \\
Fixed Network & 3.92 (0.63) & 0.052 (0.124) & 3.92 (0.45) & 0.027 (0.078) & 4.00 (0.35)& 0.040 (0.063)\\
Random Graph Model & 4.02 (0.68) & -3.03 (3.06) & 4.01 (0.51) & -2.11 (3.30) &  4.13 (0.34)& -0.98 (2.75)\\
CINet & 3.99 (0.015) & 5.49 (0.92) & 4.00 (0.012) & 5.53 (0.69) & 4.00 (0.013) & 6.52 (1.02) \\
True Value & 4.0 & 5.5 & 4.0 & 5.5 & 4.0 & 6.5\\
\bottomrule
\end{tabular}}
\label{table1}
\end{table}

\section{Case Study}

In this section, we apply our proposed method to the stroke patient transfer network data. The data was collected from all 342 non-federal California hospitals, including around 350,000 stroke patients from 2010-2020. Our goal is to evaluate the treatment effects of a specific reperfusion intervention -- EVT on patient outcomes. Specifically, the outcome is defined by the log transformed 30/60/90-day mortality rate per hospital. The treatment per hospital is a binary indicator indicating whether a hospital is capable of performing EVT or not. Access to timely interventions is critical in stroke. Patients are sometimes transferred to hospitals to access EVT that are typically available only in a subset of hospitals. We constructed the hospital network by connecting two hospitals if there was at least one patient transferred from one to the other. We only considered transfers that were initiated from emergency departments and happened within 24 hours of initial admission. 

\begin{figure}
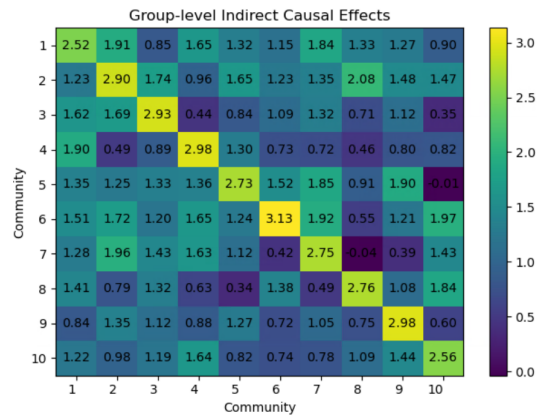

\figuresize{.3}
\figurebox{20pc}{25pc}{}[fig2]
\caption{Heatmap of posterior means of group-level indirect treatment effects on 30-day mortality.}
\label{fig2}
\end{figure}

We use all encounter data from 2019 as our running example. The final dataset consisted of 190 hospitals, with an average of 7.7 patients transferred out per sending hospital and 14.2 patients accepted per receiving hospital. We applied the community detection algorithm to the patient transfer network, which leads us to 10 non-overlapping hospital clusters. Based on the cluster labels, we calculate the direct and group-wise indirect effects of EVT capability on hospital morality rates. Results using 30-day mortality are shown in Figure~\ref{fig2}. Results using 60-day mortality and 90-day mortality are similar to results shown in Figure~\ref{fig2}. See more details in supplementary materials.

\section{Discussion}

This paper studies causal inference with an unknown interference network. We target group-level effects that summarize how treatments propagate within and across latent communities, and we propose a two-step framework for effect estimation. Our contributions are: (1) we establish identification conditions for heterogeneous group-level treatment effects; (2) we prove asymptotic properties of the maximum likelihood estimator (MLE); and (3) we develop a Bayesian framework to estimate posterior distributions of the model parameters and establish their asymptotic properties. Simulations and an application to the HCAI data show that our approach recovers heterogeneous group effects and yields interpretable summaries of interference pathways.

We focus on continuous outcomes in this study. An immediate extension is to handle discrete outcomes (e.g., binary or counts) within a generalized modeling framework. We also treat the network as static, whereas in practice the network, outcomes, and treatment assignments may evolve over time. Extending our estimands and methods to dynamic networks with time-varying treatments and outcomes—while accounting for temporal interference—remains an important direction for future work.

By explicitly modeling interference effects under latent community structure, our framework advances the causal inference literature in settings where the interference network is unknown. Theoretically, it establishes identification conditions and provides estimation strategies for heterogeneous group-level effects, thereby extending existing work on causal inference with interference to heterogeneous unobserved networks. Clinically, it offers a principled way to evaluate intervention effects where interference is pervasive—for example, assessing how the availability of endovascular thrombectomy (EVT) at one hospital improves outcomes for both its own patients and those transferred from surrounding hospitals. More broadly, the framework helps resource allocation and policy design in networked healthcare delivery, community-based interventions, and public health programs, where recognizing and quantifying interference effects is essential for improving population outcomes.

\bibliographystyle{achemso}
\bibliography{main}

\providecommand{\latin}[1]{#1}
\makeatletter
\providecommand{\doi}
  {\begingroup\let\do\@makeother\dospecials
  \catcode`\{=1 \catcode`\}=2 \doi@aux}
\providecommand{\doi@aux}[1]{\endgroup\texttt{#1}}
\makeatother
\providecommand*\mcitethebibliography{\thebibliography}
\csname @ifundefined\endcsname{endmcitethebibliography}  {\let\endmcitethebibliography\endthebibliography}{}
\begin{mcitethebibliography}{17}
\providecommand*\natexlab[1]{#1}
\providecommand*\mciteSetBstSublistMode[1]{}
\providecommand*\mciteSetBstMaxWidthForm[2]{}
\providecommand*\mciteBstWouldAddEndPuncttrue
  {\def\EndOfBibitem{\unskip.}}
\providecommand*\mciteBstWouldAddEndPunctfalse
  {\let\EndOfBibitem\relax}
\providecommand*\mciteSetBstMidEndSepPunct[3]{}
\providecommand*\mciteSetBstSublistLabelBeginEnd[3]{}
\providecommand*\EndOfBibitem{}
\mciteSetBstSublistMode{f}
\mciteSetBstMaxWidthForm{subitem}{(\alph{mcitesubitemcount})}
\mciteSetBstSublistLabelBeginEnd
  {\mcitemaxwidthsubitemform\space}
  {\relax}
  {\relax}

\bibitem[Rosenbaum(2007)]{rosenbaum2007interference}
Rosenbaum,~P.~R. Interference between units in randomized experiments. \emph{Journal of the American Statistical Association} \textbf{2007}, \emph{102}, 191--200\relax
\mciteBstWouldAddEndPuncttrue
\mciteSetBstMidEndSepPunct{\mcitedefaultmidpunct}
{\mcitedefaultendpunct}{\mcitedefaultseppunct}\relax
\EndOfBibitem
\bibitem[Hudgens and Halloran(2008)Hudgens, and Halloran]{hudgens2008toward}
Hudgens,~M.~G.; Halloran,~M.~E. Toward causal inference with interference. \emph{Journal of the american statistical association} \textbf{2008}, \emph{103}, 832--842\relax
\mciteBstWouldAddEndPuncttrue
\mciteSetBstMidEndSepPunct{\mcitedefaultmidpunct}
{\mcitedefaultendpunct}{\mcitedefaultseppunct}\relax
\EndOfBibitem
\bibitem[Tchetgen and VanderWeele(2012)Tchetgen, and VanderWeele]{tchetgen2012causal}
Tchetgen,~E. J.~T.; VanderWeele,~T.~J. On causal inference in the presence of interference. \emph{Statistical methods in medical research} \textbf{2012}, \emph{21}, 55--75\relax
\mciteBstWouldAddEndPuncttrue
\mciteSetBstMidEndSepPunct{\mcitedefaultmidpunct}
{\mcitedefaultendpunct}{\mcitedefaultseppunct}\relax
\EndOfBibitem
\bibitem[VanderWeele \latin{et~al.}(2015)VanderWeele, Tchetgen, and Halloran]{vanderweele2015interference}
VanderWeele,~T.~J.; Tchetgen,~E. J.~T.; Halloran,~M.~E. Interference and sensitivity analysis. \emph{Statistical science: a review journal of the Institute of Mathematical Statistics} \textbf{2015}, \emph{29}, 687\relax
\mciteBstWouldAddEndPuncttrue
\mciteSetBstMidEndSepPunct{\mcitedefaultmidpunct}
{\mcitedefaultendpunct}{\mcitedefaultseppunct}\relax
\EndOfBibitem
\bibitem[Manski(2013)]{manski2013identification}
Manski,~C.~F. Identification of treatment response with social interactions. \emph{The Econometrics Journal} \textbf{2013}, \emph{16}, S1--S23\relax
\mciteBstWouldAddEndPuncttrue
\mciteSetBstMidEndSepPunct{\mcitedefaultmidpunct}
{\mcitedefaultendpunct}{\mcitedefaultseppunct}\relax
\EndOfBibitem
\bibitem[Sussman and Airoldi(2017)Sussman, and Airoldi]{sussman2017elements}
Sussman,~D.~L.; Airoldi,~E.~M. Elements of estimation theory for causal effects in the presence of network interference. \emph{arXiv preprint arXiv:1702.03578} \textbf{2017}, \relax
\mciteBstWouldAddEndPunctfalse
\mciteSetBstMidEndSepPunct{\mcitedefaultmidpunct}
{}{\mcitedefaultseppunct}\relax
\EndOfBibitem
\bibitem[Athey \latin{et~al.}(2018)Athey, Eckles, and Imbens]{athey2018exact}
Athey,~S.; Eckles,~D.; Imbens,~G.~W. Exact p-values for network interference. \emph{Journal of the American Statistical Association} \textbf{2018}, \emph{113}, 230--240\relax
\mciteBstWouldAddEndPuncttrue
\mciteSetBstMidEndSepPunct{\mcitedefaultmidpunct}
{\mcitedefaultendpunct}{\mcitedefaultseppunct}\relax
\EndOfBibitem
\bibitem[Hu \latin{et~al.}(2022)Hu, Li, and Wager]{hu2022average}
Hu,~Y.; Li,~S.; Wager,~S. Average direct and indirect causal effects under interference. \emph{Biometrika} \textbf{2022}, \emph{109}, 1165--1172\relax
\mciteBstWouldAddEndPuncttrue
\mciteSetBstMidEndSepPunct{\mcitedefaultmidpunct}
{\mcitedefaultendpunct}{\mcitedefaultseppunct}\relax
\EndOfBibitem
\bibitem[Ogburn \latin{et~al.}(2024)Ogburn, Sofrygin, Diaz, and Van~der Laan]{ogburn2024causal}
Ogburn,~E.~L.; Sofrygin,~O.; Diaz,~I.; Van~der Laan,~M.~J. Causal inference for social network data. \emph{Journal of the American Statistical Association} \textbf{2024}, \emph{119}, 597--611\relax
\mciteBstWouldAddEndPuncttrue
\mciteSetBstMidEndSepPunct{\mcitedefaultmidpunct}
{\mcitedefaultendpunct}{\mcitedefaultseppunct}\relax
\EndOfBibitem
\bibitem[Zhang and Imai(2023)Zhang, and Imai]{zhang2023individualized}
Zhang,~Y.; Imai,~K. Individualized policy evaluation and learning under clustered network interference. \emph{arXiv preprint arXiv:2311.02467} \textbf{2023}, \relax
\mciteBstWouldAddEndPunctfalse
\mciteSetBstMidEndSepPunct{\mcitedefaultmidpunct}
{}{\mcitedefaultseppunct}\relax
\EndOfBibitem
\bibitem[Leung(2022)]{leung2022causal}
Leung,~M.~P. Causal inference under approximate neighborhood interference. \emph{Econometrica} \textbf{2022}, \emph{90}, 267--293\relax
\mciteBstWouldAddEndPuncttrue
\mciteSetBstMidEndSepPunct{\mcitedefaultmidpunct}
{\mcitedefaultendpunct}{\mcitedefaultseppunct}\relax
\EndOfBibitem
\bibitem[S{\"a}vje(2024)]{savje2024causal}
S{\"a}vje,~F. Causal inference with misspecified exposure mappings: separating definitions and assumptions. \emph{Biometrika} \textbf{2024}, \emph{111}, 1--15\relax
\mciteBstWouldAddEndPuncttrue
\mciteSetBstMidEndSepPunct{\mcitedefaultmidpunct}
{\mcitedefaultendpunct}{\mcitedefaultseppunct}\relax
\EndOfBibitem
\bibitem[Li \latin{et~al.}(2021)Li, Sussman, and Kolaczyk]{li2021causal}
Li,~W.; Sussman,~D.~L.; Kolaczyk,~E.~D. Causal inference under network interference with noise. \emph{arXiv preprint arXiv:2105.04518} \textbf{2021}, \relax
\mciteBstWouldAddEndPunctfalse
\mciteSetBstMidEndSepPunct{\mcitedefaultmidpunct}
{}{\mcitedefaultseppunct}\relax
\EndOfBibitem
\bibitem[Lewbel \latin{et~al.}(2024)Lewbel, Qu, and Tang]{lewbel2024ignoring}
Lewbel,~A.; Qu,~X.; Tang,~X. Ignoring measurement errors in social networks. \emph{The Econometrics Journal} \textbf{2024}, \emph{27}, 171--187\relax
\mciteBstWouldAddEndPuncttrue
\mciteSetBstMidEndSepPunct{\mcitedefaultmidpunct}
{\mcitedefaultendpunct}{\mcitedefaultseppunct}\relax
\EndOfBibitem
\bibitem[Li and Wager(2022)Li, and Wager]{li2022random}
Li,~S.; Wager,~S. Random graph asymptotics for treatment effect estimation under network interference. \emph{The Annals of Statistics} \textbf{2022}, \emph{50}, 2334--2358\relax
\mciteBstWouldAddEndPuncttrue
\mciteSetBstMidEndSepPunct{\mcitedefaultmidpunct}
{\mcitedefaultendpunct}{\mcitedefaultseppunct}\relax
\EndOfBibitem
\bibitem[Zhao \latin{et~al.}(2012)Zhao, Levina, and Zhu]{zhao2012consistency}
Zhao,~Y.; Levina,~E.; Zhu,~J. Consistency of community detection in networks under degree-corrected stochastic block models. \textbf{2012}, \relax
\mciteBstWouldAddEndPunctfalse
\mciteSetBstMidEndSepPunct{\mcitedefaultmidpunct}
{}{\mcitedefaultseppunct}\relax
\EndOfBibitem
\bibitem[Horvitz and Thompson(1952)Horvitz, and Thompson]{horvitz1952generalization}
Horvitz,~D.~G.; Thompson,~D.~J. A generalization of sampling without replacement from a finite universe. \emph{Journal of the American statistical Association} \textbf{1952}, \emph{47}, 663--685\relax
\mciteBstWouldAddEndPuncttrue
\mciteSetBstMidEndSepPunct{\mcitedefaultmidpunct}
{\mcitedefaultendpunct}{\mcitedefaultseppunct}\relax
\EndOfBibitem
\end{mcitethebibliography}


\end{document}